\documentclass[prd,aps,showpacs,twocolumn,preprintnumbers,nofootinbib,amsmath,amssymb,superscriptaddress,longbibliography]{revtex4-2}
\usepackage{xcolor}
\usepackage{graphicx}

\usepackage{xcolor}
\usepackage{graphicx}

\newcommand{\beqn}{\begin{eqnarray}}
\newcommand{\eeqn}{\end{eqnarray}}
\newcommand{\beqs}{\begin{subequations}}
\newcommand{\eeqs}{\end{subequations}\\[-2mm]\noindent}
\newcommand{\eq}[1]{(\ref{#1})}

\newcommand{\bs}{\boldsymbol}
\newcommand{\avr}[1]{{\left\langle #1 \right\rangle}}

\makeatletter
\newcommand{\ostar}{\mathbin{\mathpalette\make@circled*}}
\newcommand{\make@circled}[2]{%
  \ooalign{$\m@th#1\smallbigcirc{#1}$\cr\hidewidth$\m@th#1#2$\hidewidth\cr}%
}
\newcommand{\smallbigcirc}[1]{%
  \vcenter{\hbox{\scalebox{0.6}{$\m@th#1\bigcirc$}}}%
}
\makeatother

\def\bbbone{{\mathchoice {\rm 1\mskip-4mu l} {\rm 1\mskip-4mu l} {\rm 1\mskip-4.5mu l} {\rm 1\mskip-5mu l}}}

\definecolor{brickred}{rgb}{0.8, 0.25, 0.33}

\allowdisplaybreaks

\begin{document}

\bibliographystyle{apsrev4-1}

\title{Ising Model on the Fibonacci Sphere}

\author{A. S. Pochinok}
\affiliation{Pacific Quantum Center, Far Eastern Federal University, 690950 Vladivostok, Russia}
\author{A. V. Molochkov}
\affiliation{Pacific Quantum Center, Far Eastern Federal University, 690950 Vladivostok, Russia}
\author{M. N. Chernodub}
\affiliation{Institut Denis Poisson UMR 7013, Universit\'e de Tours, 37200 Tours, France}

\begin{abstract}
We formulate the ferromagnetic Ising model on a two-dimensional sphere using the Delaunay triangulation of the Fibonacci covering. The Fibonacci approach generates a uniform isotropic covering of the sphere with approximately equal-area triangles, thus potentially supporting a smooth thermodynamic limit. In the absence of magnetic field, the model exhibits a spontaneous magnetization phase transition at a critical temperature that depends on the connectivity properties of the underlying lattice. While in the standard triangular lattice, every site is connected to 6 neighboring sites, the triangulated Fibonacci lattice of the curved surface contains a substantial density of the 5- and 7-vertices. As the number of sites in the Fibonacci sphere increases, the triangular cover of the sphere experiences a series of singular transitions that reflect a sudden change in its connectivity properties. These changes substantially influence the statistical features of the system leading to a series of first-order-like discontinuities as the radius of the sphere increases. We found that the Ising model on a uniform, Fibonacci-triangulated sphere in a large-radius limit possesses the phase transition at the critical temperature $T^{\large\ostar}_c \simeq 3.33(3) J$, which is slightly lower than the thermodynamic result for an equilaterally triangulated planar lattice. This mismatch is a memory effect: the planar Fibonacci lattice remembers its origin from the curved space.
\end{abstract}

\date{\today}

\maketitle

\section{Introduction}

The two-dimensional Ising model is one of the simplest statistical models that describe ferromagnetism in planar lattices associated with $2d$ crystalline atomic systems~\cite{LL5}. The spin variable $S_{\bs x}$ represents the magnetic moment of an atom at the site $\bs x$. This variable can take two values, $S_{\bs x} = + 1$ or $S_{\bs x} = - 1$, associated, respectively, with up or down orientations of the magnetic moment with respect to the atomic crystal plane. The Hamiltonian of the system is:
\begin{equation}
H = - J \sum\limits_{<{\bs x}, {\bs y}>} S_{\bs x} S_{\bs y} - h \sum\limits_{\bs x} S_{\bs x}\,,
\label{eq_H} 
\end{equation} 
where $J$ is the interaction energy and $h$ is the external magnetic field. The summation in Eq.~\eq{eq_H} is going over the pairs  $<{\bs x}, {\bs y}>$ of nearest-neighbor sites.

At low temperatures, the Ising model~\eq{eq_H} with positive spin coupling $J>0$ possesses a ferromagnetic phase characterized by the ordered (co-aligned) spins of the same sign. At high temperatures, the spins reside in a disordered, paramagnetic phase. The model describes a class of natural ferromagnetic systems. 

In the absence of the background magnetic field, $h = 0$, the model~\eq{eq_H} possesses the global ${\mathbb Z}_2$ symmetry, which corresponds to the invariance of the $h = 0$ Hamiltonian with respect to a simultaneous flip of all spins, $S_{\bs x} \to - S_{\bs x}$. The ordered ferromagnetic ground state spontaneously breaks this symmetry, ${\mathbb Z}_2 \to \bbbone$. The ordered and disordered phases are separated by a second-order thermodynamic phase transition at a critical temperature $T_c$. The transition temperature and the associated critical exponents depend on the type (connectivity properties) of the underlying lattice. For example, the critical temperature on the square lattice~\cite{Kramers:1941kn,baxter2016exactly} 
\beqn
\frac{T^\Box_c}{J} = \frac{2}{\ln \left(1 + \sqrt{2}\right)} \simeq 2.2692\,,
\label{eq_T_square}
\eeqn
is substantially lower than the transition temperature at the equilateral triangular lattice~\cite{baxter2016exactly}:
\beqn
\frac{T^\triangle_c}{J} = \frac{4}{\ln 3} \simeq 3.6410\,,
\label{eq_T_triangle}
\eeqn
(here, we work in the units $k_B = 1$). For the first time, the Ising model was considered on a triangular lattice in works~\cite{Newell_1950, Wannier_1950}. A detailed calculation of the critical parameters of the phase transition of the model has been implemented in Ref.~\cite{Zhi_Huan_2009}.

At a negative spin coupling, $J<0$, the nearest spins tend to anti-align, thus forming an (ordered) antiferromagnetic phase. The structure of the ground state depends on the connectivity features of the underlying lattice. For bipartite lattices (the ones that can be divided into two sub-lattices with sites from one sub-lattice interacting with sites from the other sub-lattice only), the ground state corresponds to the opposite orientation of the spins at fully polarized sub-lattices. The ground state structure is less straightforward for non-bipartite lattices (for example, for a triangular lattice). The antiferromagnetic Ising model possesses plenty of interesting phenomena associated with frustrated ground states~\cite{collins1997review}. While in our study, we concentrate on the ferromagnetic sector of the phase diagram with $J>0$, the connectivity of the lattice will also play an essential role, as we will see later (in our paper, we consider non-bipartite lattices). 

We investigate the phase diagram of the Ising model in the spherical topology. To tackle this problem, we triangulate the spherical surface via its covering by the Fibonacci lattice. The Fibonacci sphere is characterized by a uniform distribution of points over the whole area without unnatural inhomogeneities that emerge, for example, around the poles in a naive latitude–longitude triangulation in a spherical coordinate system~\cite{gonzalez2010measurement}. The Fibonacci sphere has already proven itself to be helpful in magnetic resonance modeling~\cite{Saliba_2019}, image processing~\cite{Marques_2013}, climate models \cite{Swinbank_2006}, and others. Moreover, in lattice spin (gauge) models, the Fibonacci covering has potentially a well-defined thermodynamic limit due to approximately equal-area regions between three adjacent sites (triangular plaquettes). This fact makes it attractive for formulating problems of lattice quantum field theory in spaces with a spherical topology and, potentially, at other surfaces (the Fibonacci lattice has been recently employed in the investigation of the XY model in Ref.~\cite{Song2021}). In Section~\ref{sec:triangulation}, we discuss the generation of a Fibonacci lattice on the sphere and its triangulation via the Delaunay method.

The thermodynamic properties of the model on the Fibonacci sphere are discussed in Section~\ref{sec:thermodynamics}. We consider various quantities, such as magnetization and energy, as well as the corresponding susceptibilities. We determine the position of the phase transition in the absence of the magnetic field and investigate the thermodynamic limit of the model. The thermodynamic limit corresponds to the infinite number of points which, for a uniformly triangulated sphere, implies the limit of the sphere of an infinite radius, $R \to \infty$.

The infinite-radius limit of the Fibonacci sphere corresponds to a locally flat surface of an infinite area. Naively, one could expect the properties of the Ising model formulated on the $R \to \infty$ Fibonacci sphere and the infinite-area triangular planar lattice should be the same. We show that, unexpectedly, this expectation is not realized since, even for a thermodynamically substantial number of sites, the Fibonacci lattice contains a significant density of points with abnormal 5-vertex or 7-vertex connectivities that influence the thermodynamic properties of the system. In other words, the Fibonacci lattice in the flat, infinite-radius limit still ``remembers'' its spherical origin and differs from the standard equilateral lattice. 

Our conclusions are summarized in Section~\ref{sec:conclusions}.

\section{Triangulated Fibonacci sphere}
\label{sec:triangulation}

\subsection{Fibonacci lattice}

The Fibonacci lattice is characterized by an almost uniform and isotropic distribution of the lattice sites. These properties play an essential role in the triangulation of the sphere, which is challenging to triangulate uniformly otherwise~\cite{gonzalez2010measurement,Saff_1997}. 

A Fibonacci number $F_i$ is a part of the infinite sequence in which every term, from the third term onward ($i \geqslant 3$), is the sum of the previous two terms, $F_{i} = F_{i-1} + F_{i-2}$. The first few numbers in the sequence are as follows:
\beqn
\{ 0, 1, 1, 2, 3, 5, 8, 13, 21, \dots\}\,.
\label{eq:F:numbers}
\eeqn
The ratio of two adjacent numbers gives, in the large-number limit $i \to \infty$, the golden ratio
\beqn
g = \lim_{i \to \infty} \frac{F_{i+1}}{F_i} = \frac{1 + \sqrt{5}}{2} \approx 1.618 \approx 137.5^\circ\,,
\label{eq:golden:ratio}
\eeqn
defined by the relationship $g = 1 + 1/g$.

\begin{figure}[!htb]
\includegraphics[width=0.4\textwidth]{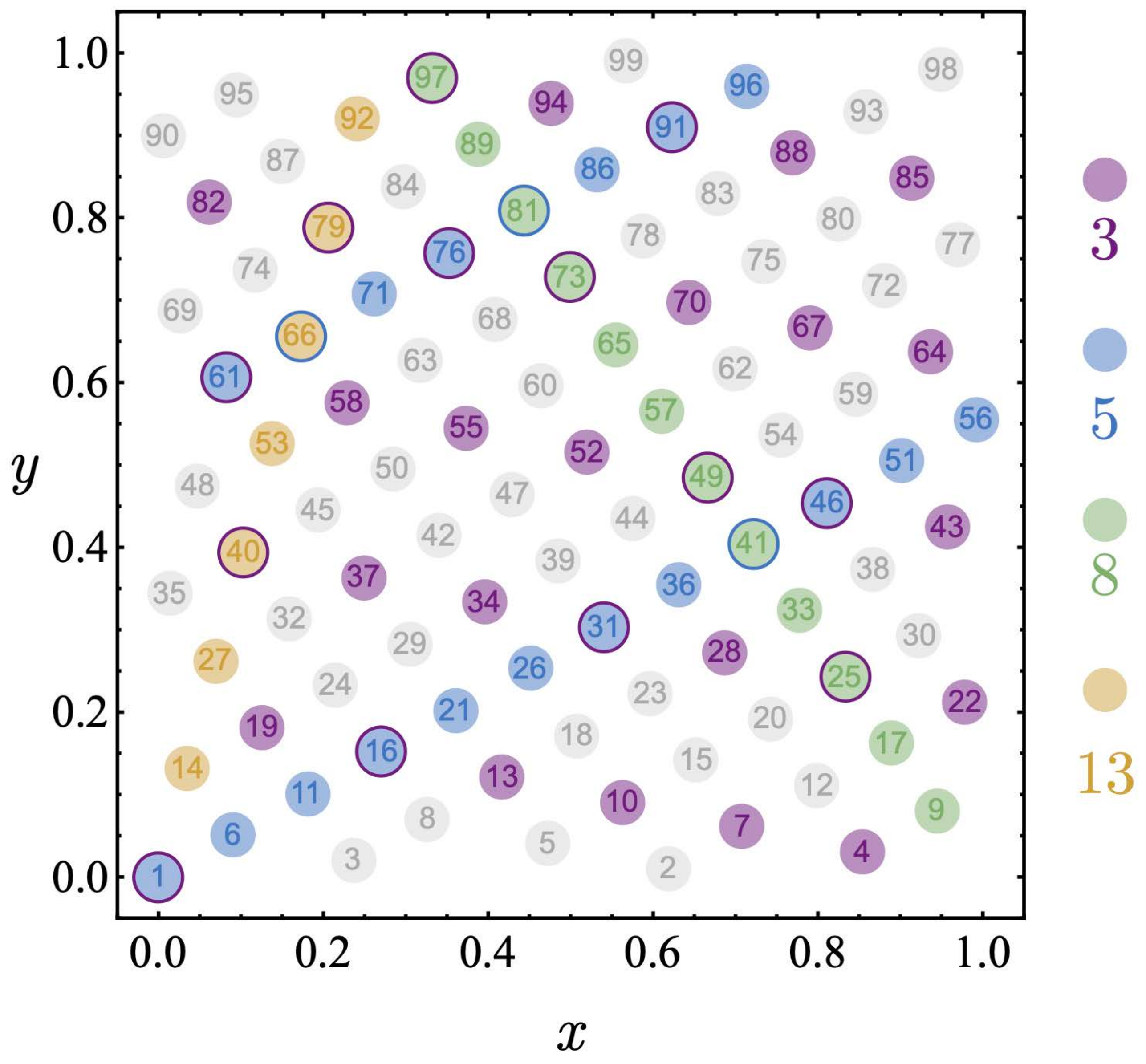}
\caption{A flat Fibonacci lattice made of 99 sites using sche\-me~\eq{eq:square}. Ordinal numbering is used. The sequences with the Fibonacci numbers 3, 5, 8, and 13 are shown in colors.}
\label{fig:Fibonacci:101}
\end{figure}

The structure of the Fibonacci lattice is most straightforward to understand in a flat space. The Fibonacci covering of a unit square by a set of points is as follows~\cite{Niederreiter_1994}:
\beqn
\label{eq:square} 
x_k = \left\{\frac{k}{g}\,\right\}, \qquad y_k = \frac{k}{n}\,,
\eeqn
where the integer number $k=1,\dots, n$ labels the sites with the coordinates ${\bs x} = (x_k, y_k)$ and $\left\{\dots\right\}$ stands for the fractional part of a number. The resulting uniform distribution of the sites over a unit square is shown in Fig.~\ref{fig:Fibonacci:101}.

It is easy to conclude from Fig.~\ref{fig:Fibonacci:101} that the distribution of the points is reasonably uniform. Moreover, the sites are structured in various straight lines, inclined clockwise and counterclockwise, with the difference between the adjacent elements given by a Fibonacci number~\eq{eq:F:numbers}. The latter property gives the name of the whole construction, called ``the Fibonacci lattice''. Notice that, in general, each site is a part of more than one structured line with distinct lines labeled by a fixed Fibonacci number. Some of these lines are easy to spot in Fig.~\ref{fig:Fibonacci:101}.

A similar construction on a disk is called the Fibonacci spiral. This structure is obtained by imposing a grid on a disk using the following parametrization of the points in the polar coordinate system~\cite{Swinbank_2006, Vogel_1979}:
\beqs
\beqn
x & = & r \cos\varphi\,,\qquad \qquad y = r \sin\varphi\,,\\
r & = & r_0 \sqrt{k -\frac{1}{2}}\,,\qquad  \varphi =\frac{2\pi k}{g},
\eeqn
\label{eq:spiral}
\eeqs
where $r_0$ is an arbitrary scale parameter and $k=1,2,\dots$ is an integer number that labels the lattice sites.

\begin{figure}[!htb]
\includegraphics[width=0.475\textwidth]{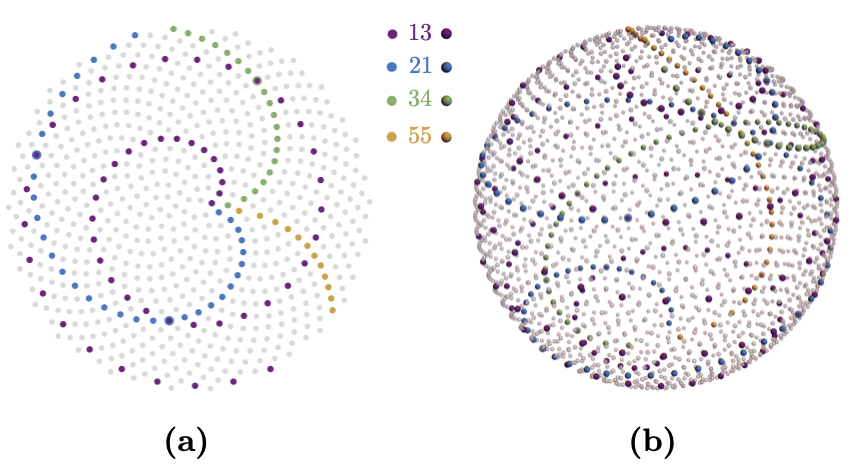}
\caption{(a) Fibonacci spiral a disk with 700 sites constructed with scheme~\eq{eq:spiral}. In general, the spiral structure of the Fibonacci lattice is not evident since the distance between the points of the sequence is bigger than the distance between the geometrically closest points. Here, every 13's, 21's, 34's, and 55's sites, corresponding to subsequent Fibonacci numbers, are highlighted in color to visualize the spiral structure. (b) The Fibonacci lattice on a sphere made with 2001 sites using construction~\eq{eq:sphere}.}
\label{fig_spirals}
\end{figure}

The Fibonacci spiral is shown in Fig.~\ref{fig_spirals}(a). This structure can be represented in two sets of spiral arcs labeled by the Fibonacci number~\eq{eq:F:numbers}, with the latter representing the distance between the consecutive points. The construction is pretty similar to the case of the disk with the difference given by the curved nature of the spirals that are often called ``dominant''~\cite{Swinbank_2006}. Some sets of spirals are twisted clockwise, while in the other set, they are twisted in an opposite, counterclockwise direction. With an increase in the lattice radius (total number of points), the number of the dominant spirals increases. One can also show that the Fibonacci lattice is, in fact, the result of a sequential arrangement of elements in the form of a single, ``generative'' spiral~\cite{Bravais1837}. The angular distance between the elements of the generative spiral corresponds to the golden ratio~\eq{eq:golden:ratio}. 
Since the golden ratio is the most irrational number~\cite{weisstein2002crc}, the appearance of this number in the construction of the Fibonacci spiral is an essential factor that serves to optimize the packing efficiency in avoiding periodicities or quasi-periodicities of the elements.

In order to obtain a uniform Fibonacci covering of a sphere, it is enough to project a flat uniform lattice on a disk (the Fibonacci spiral) onto a sphere to preserve equal areas between adjacent sites:
\beqs
\beqn
x & = & r \cos\vartheta\sin\varphi,\quad y = r \sin\vartheta\sin\varphi,\quad z = r \cos\varphi\,, \quad \\
\varphi & = & \frac{2\pi k}{g};\qquad  \vartheta = \arccos{\frac{k}{n}}\,.
\eeqn
\label{eq:sphere}
\eeqs
The methodology of construction and review of some properties of the Fibonacci sphere can be found in Refs.~\cite{Gonz_lez_2009, Hannay_2004, Keinert_2015}.  

\subsection{Delaunay triangulation of a sphere}

Triangulation is the process of the construction of a graph from triangular elements. For curved two-dimensional manifolds, triangulation serves as a way to find an approximation of the smooth curved surface by a set of triangles. Furthermore, triangulation is helpful in computer modeling as it allows discretizing continuous curved objects. We will also use triangulation to discretize the sphere and define the Ising model on a spherical surface. 

In order to obtain a well-defined and self-consistent thermodynamic limit of the statistical model on a curved surface, we need to approximate this surface as a collection of a connected set of triangles of approximately equal areas and shapes. In other words, the triangulation should be globally and locally uniform regarding the spatial distribution of the triangles and the local shapes of their elements.

\begin{figure}[!htb]
\begin{center}
\begin{tabular}{cc}
\includegraphics[width=0.2\textwidth]{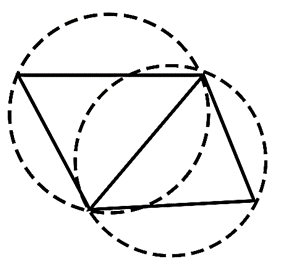} \qquad \qquad & 
\includegraphics[width=0.2\textwidth]{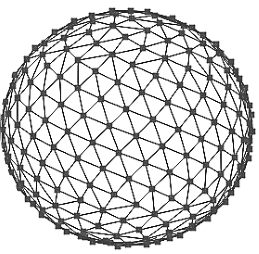} \\[3mm]
(a) & (b) 
\end{tabular}
\end{center}
\caption{Delaunay triangulation (a) for four points on a plane and (b) on a sphere of a lattice with 100 sites.}
\label{fig_Delaunay}
\end{figure}

We have already constructed the uniform distribution of points over a sphere using the Fibonacci construction. The next step is to triangulate the sphere using these points as the vertices of the triangles. For a given set of points, triangulation is generally not a unique procedure since a general set of points can be connected with a set of disjoint lines in more than one way. For example, four points shown in Fig.~\ref{fig_Delaunay}(a) could be triangulated in two different ways using triangles of two different shapes. 

In practice, a smooth, physically reasonable triangulation is achieved with the help of the Delaunay triangulation~\cite{Delaunay1934, Skvortsov2002}. The rule determines the Delaunay triangulation of a set of points that a circle circumscribed around any triangle does not contain any point from this set in the interior of the circle. An example of the Delaunay triangulation, along with the circumscribed circles, is shown in Fig.~\ref{fig_Delaunay}(a). There are two ways to make the triangulation, and only one of those, shown in the figure, satisfies the Delaunay criterion. One of the advantages of the Delaunay triangulation is that it maximizes the minimal angle in every triangle. The Delaunay method allows obtaining the triangles closest in their shapes and areas for a uniform collection of points.

For a particular case of a sphere, the Delaunay condition can be reformulated as a requirement that a plane that crosses three sites of any triangle should contain the remaining set of points from the same triangulation either above or below the plane. In other words, each such plane separates the space into two parts, with one part containing all other points of the set while the other part possessing no points from the same set. The illustration of the Delaunay triangulation of the sphere is shown in Fig.~\ref{fig_Delaunay}(b). The procedure details, defined uniquely, are presented in Appendix~\ref{app:Delaunay}. The statistical properties of the Delaunay triangulation of the Fibonacci sphere are discussed in Appendix~\ref{app_statistics}. These triangulation features will be important for the thermodynamic limit, which we discuss in the next section. 

\section{Ising model on Fibonacci sphere}
\label{sec:thermodynamics}

\subsection{Energy and magnetization}

After constructing the triangulated Fibonacci sphere, we define the Ising model with the Hamiltonian~\eq{eq_H}. While the triangles in our model are not equilateral, we consider the constant couplings, which do not depend on the length of the links. In this sense, our construction is similar to the implementation of the spin XY model on the Fibonacci sphere~\cite{Song2021}. To simplify notations, we consider dimensionful variables in units of the coupling $J=1$.

First, we studied the basic properties of the Ising model on small Fibonacci spheres to ensure that our understanding of the model was correct. To this end, we calculated the mean energy as the expectation value of the Hamiltonian, $E = \avr{H}$, and the total magnetization of the system:
\begin{equation}
 M =  \sum\limits_{{\bs x}} S_{\bs x}\,.
 \label{eq_M}
\end{equation} 
We used the Monte Carlo method to generate configurations using the standard Metropolis algorithm. The numerical results on the Fibonacci lattice with $1000$ sites shown below were calculated using $2 \cdot 10^6$ Monte Carlo steps. 

\begin{figure}[!htb]
\includegraphics[width=0.45\textwidth]{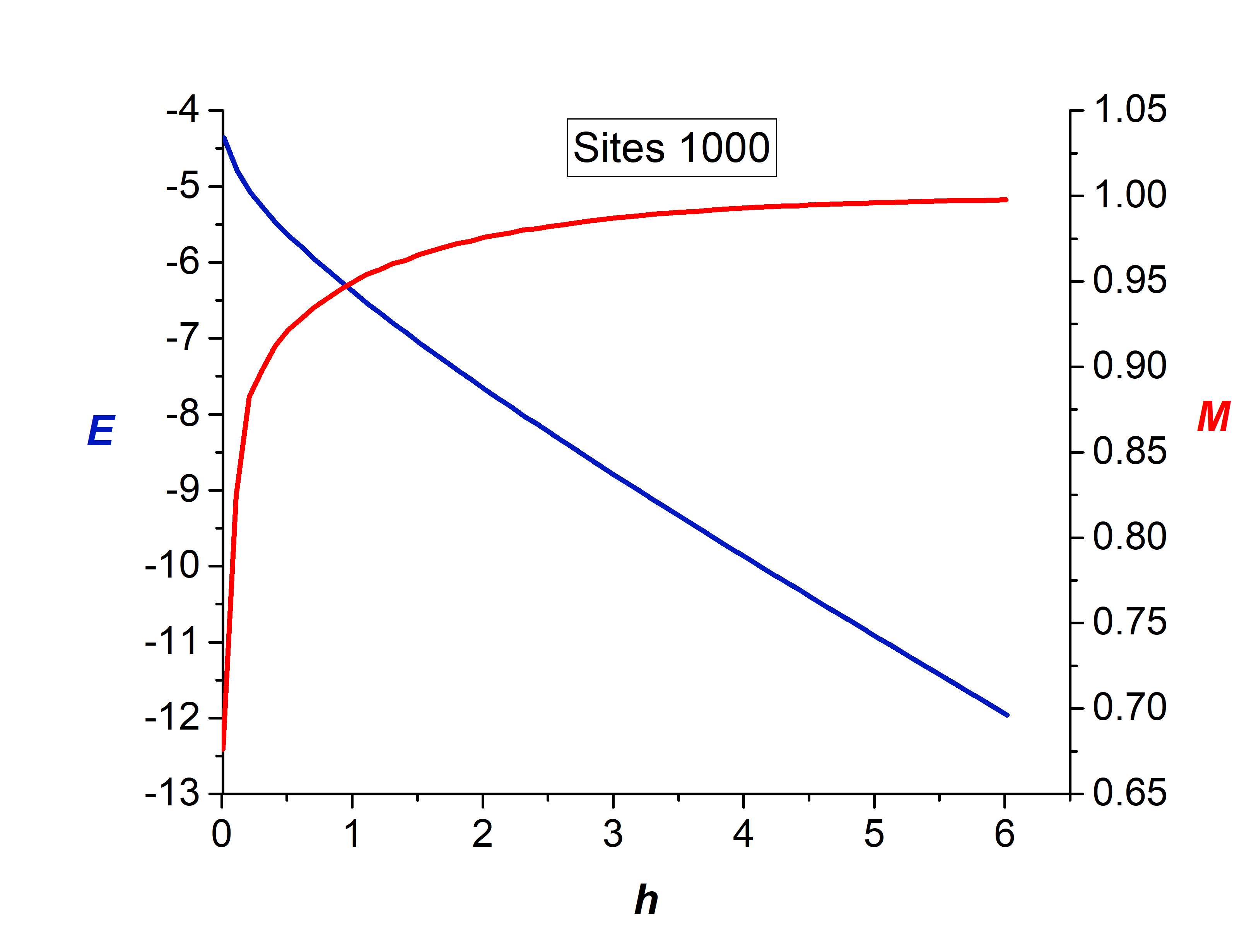}
\caption{Energy (the blue color with the axis on the left) and magnetization (the red color with the axis on the right) of the system on the Fibonacci sphere with $N = 1000$ sites as the function of the external magnetic field $h$ at fixed temperature $T = 3.5$.}
\label{fig_E_M_vs_h}
\end{figure}

In Fig.~\ref{fig_E_M_vs_h}, we show the energy and magnetization of the system as the function of the external magnetic field, $h \neq 0$. At a nonvanishing magnetic field, the spin-reflection ${\mathbb Z}_2$ symmetry is broken as the spins prefer to polarize along the direction of the magnetic field. Therefore, no critical behavior associated with the spontaneous symmetry breaking occurs at $h \neq 0$, and the transition is characterized by a smooth crossover. With increasing magnetic field $h$, the magnetization $M$ quickly reaches the saturation point (with all spins aligned along the background magnetic field) while the mean energy $E$ smoothly decreases due to ferromagnetic coupling. 

\begin{figure}[!htb]
\includegraphics[width=0.45\textwidth]{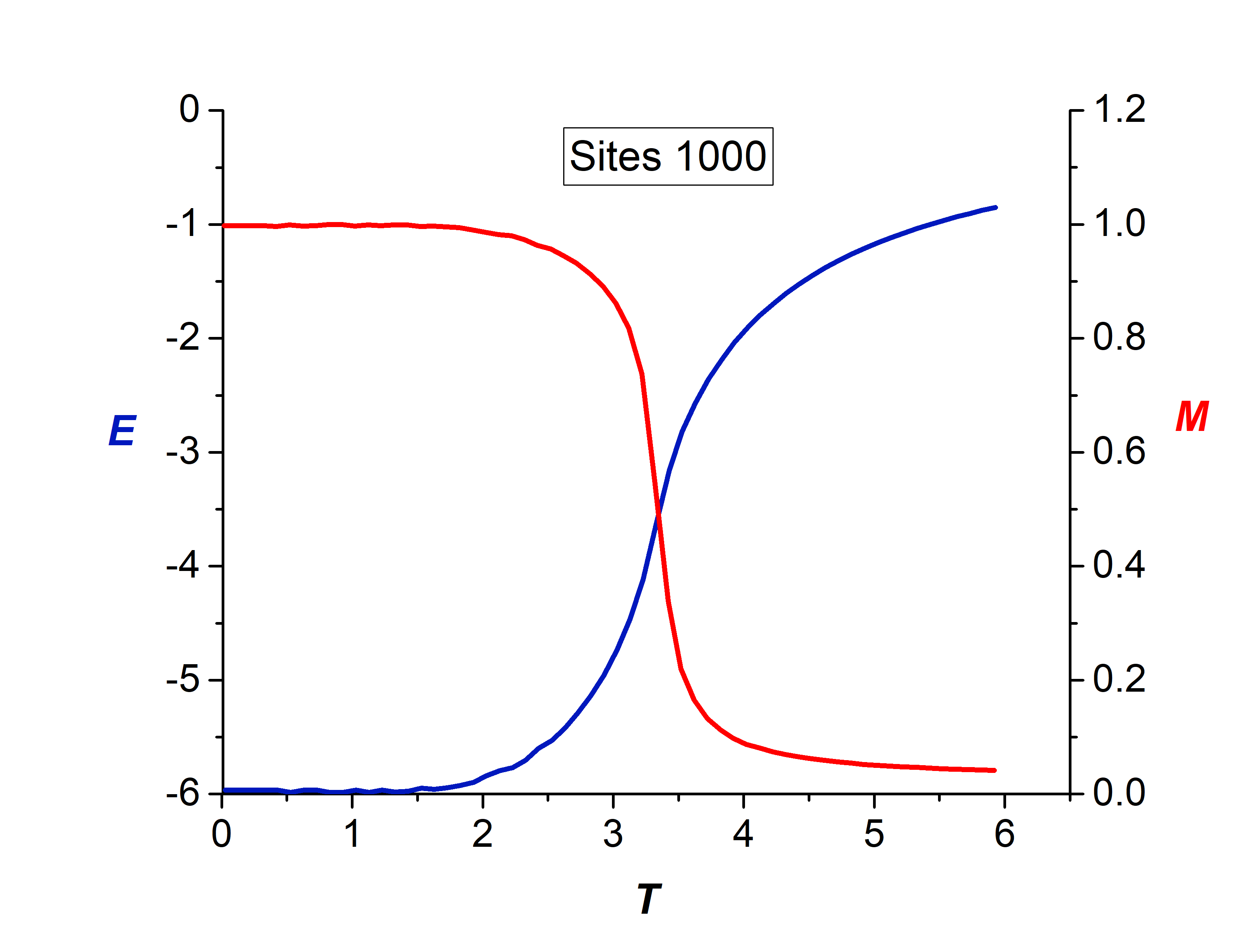}
\caption{The same as in Fig.~\ref{fig_E_M_vs_h} but as functions of temperature $T$ at vanishing external magnetic field, $h = 0$.}
\label{fig_E_M_vs_T}
\end{figure}

At vanishing magnetic field, $h = 0$, the behavior is more nontrivial as the system in thermodynamic limit experiences a phase transition at varying temperatures. On a finite lattice, the energy and magnetization exhibit a pseudo-critical behavior signaling a close presence of a thermodynamic singularity. In Fig.~\ref{fig_E_M_vs_T}, we show these quantities as functions of temperature at $h=0$. 

Both figures \ref{fig_E_M_vs_h} and \ref{fig_E_M_vs_T} show a standard behavior for the Fibonacci lattices, which is qualitatively similar to the properties of the Ising model on flat lattices~\cite{baxter2016exactly}. These features encourage us to search for the critical temperature in the thermodynamic limit similar to the standard methods used on the planar lattices. In the rest of the paper, we work at vanishing magnetic field, $h=0$. 

\subsection{Locating transition point}

An accurate way to estimate the location of the thermodynamic phase transition utilizes a Binder cumulant, a higher-order generalization of susceptibility. We use the 4th-order cumulant, $U \equiv U_4$, expressed via the magnetization~\eq{eq_M}:
\begin{equation}
U(T, N) = 1 - \frac{\langle M^4(T, N)\rangle}{3\langle M^2(T, N) \rangle^2}, 
\label{eq_CB} 
\end{equation} 
where $N$ is the number of lattice sites. The intersection of the Binder curves for different lattice sizes determines the position of phase transition in the thermodynamic limit.

\begin{figure}[!htb]
\includegraphics[width=0.4\textwidth]{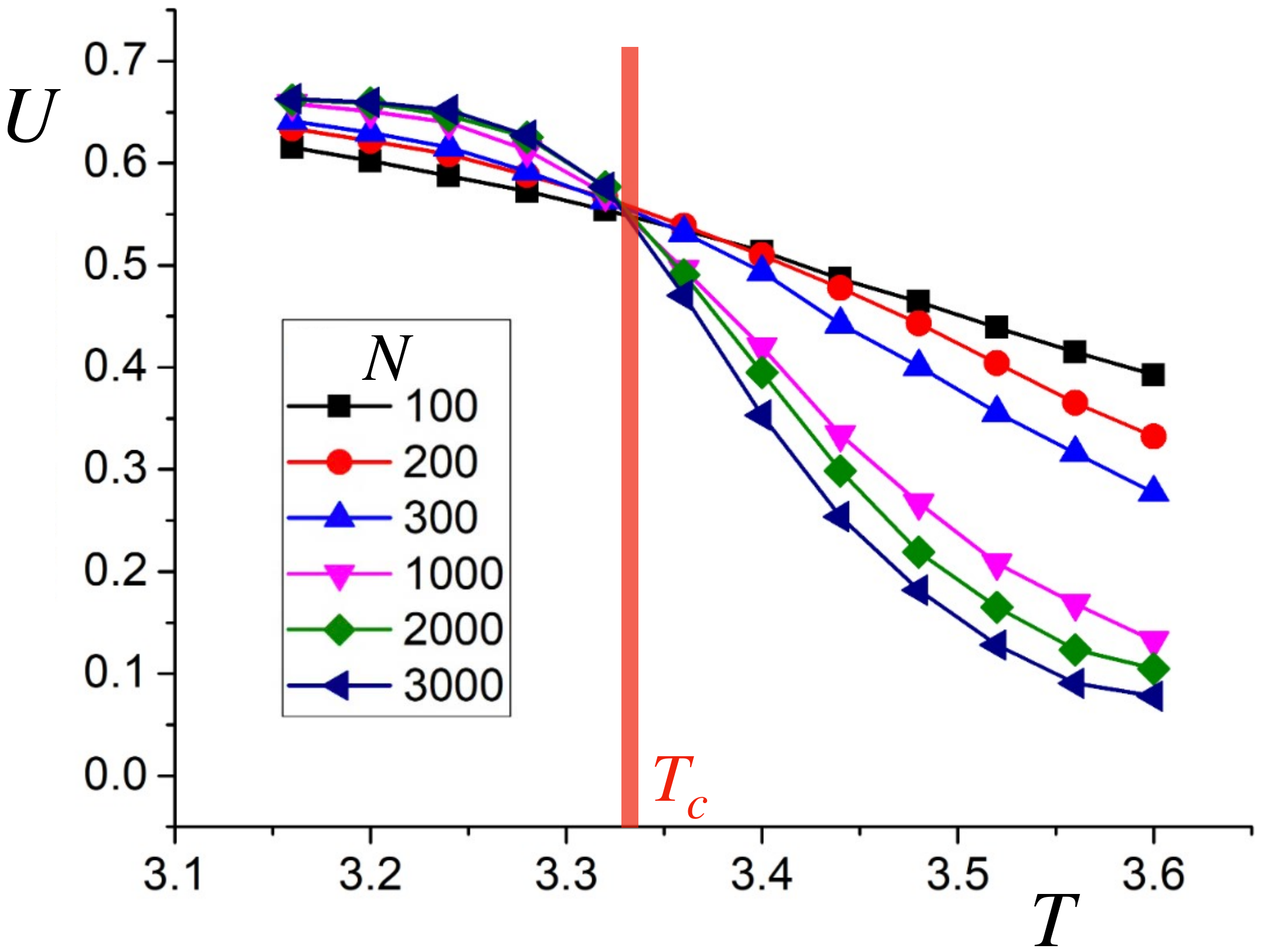}
\caption{The Binder cumulant~\eq{eq_CB} as the function of temperature $T$ at the Fibonacci lattices with different numbers of the lattice sites $N$.}
\label{fig_Binder}
\end{figure}

In Fig.~\ref{fig_Binder}, we show the Binder cumulant~\eq{eq_CB}, calculated with the number of Monte Carlo steps of the order of $10^6$, for two sets of lattices: with the rare filling (for $N = 100, 200, 300$ sites) and a denser filling (with $N = 1000, 2000, 3000$ sites). While the numerical data in the figure show a seemingly good crossing of all these curves, it does not allow us to determine the phase transition point with good precision. Surprisingly, an increase in the accuracy of Monte Carlo simulations does not help. The phase transition temperature approximately corresponds to $T_c \approx 3.35$ for the small filling of the sphere and $T_c \approx 3.32$ for the denser filling, respectively. Thus, we arrive at a contradiction that shows some inconsistent behavior of the system when we approach the thermodynamic limit.

\begin{figure}[!htb]
\includegraphics[width=0.45\textwidth]{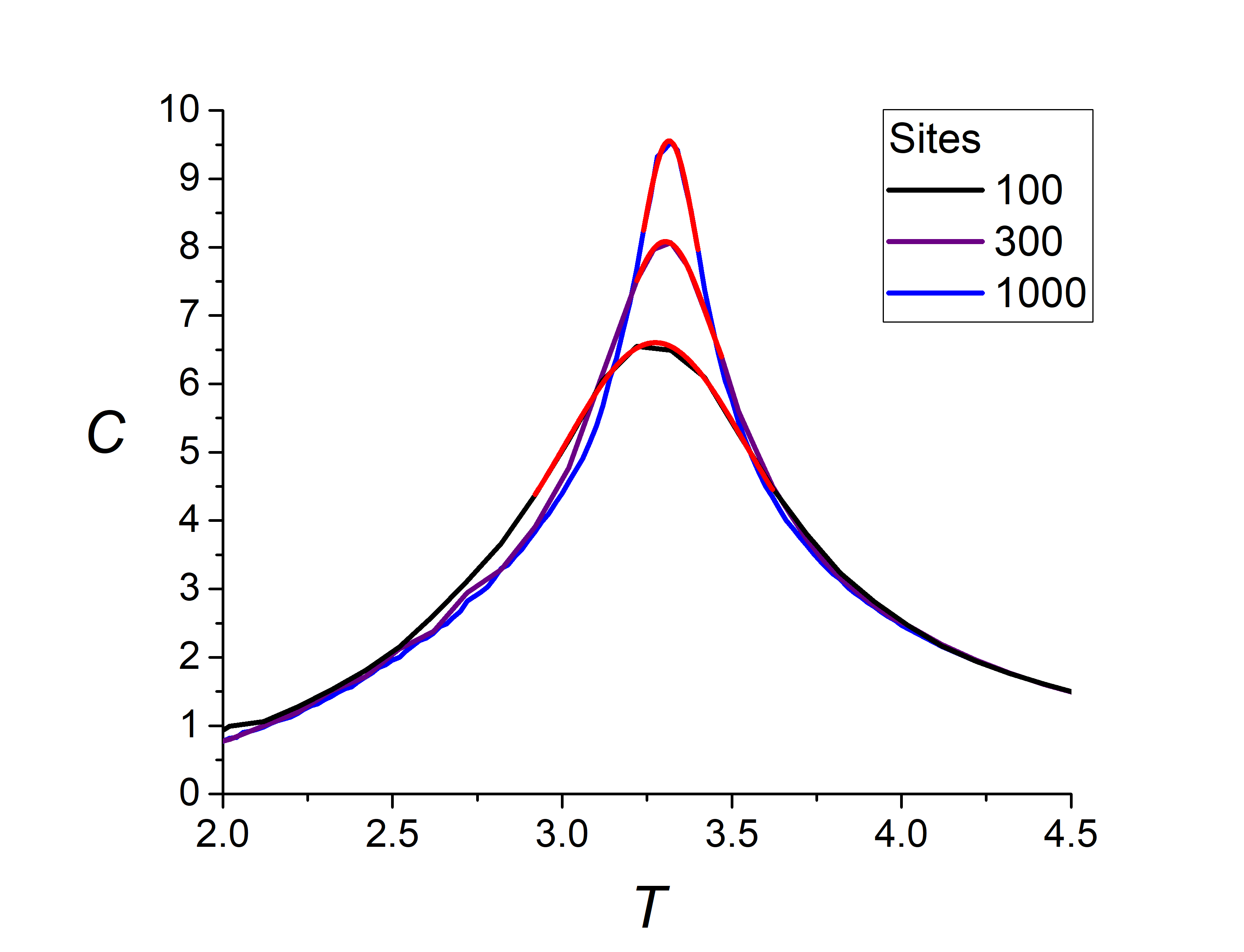}\\
\includegraphics[width=0.45\textwidth]{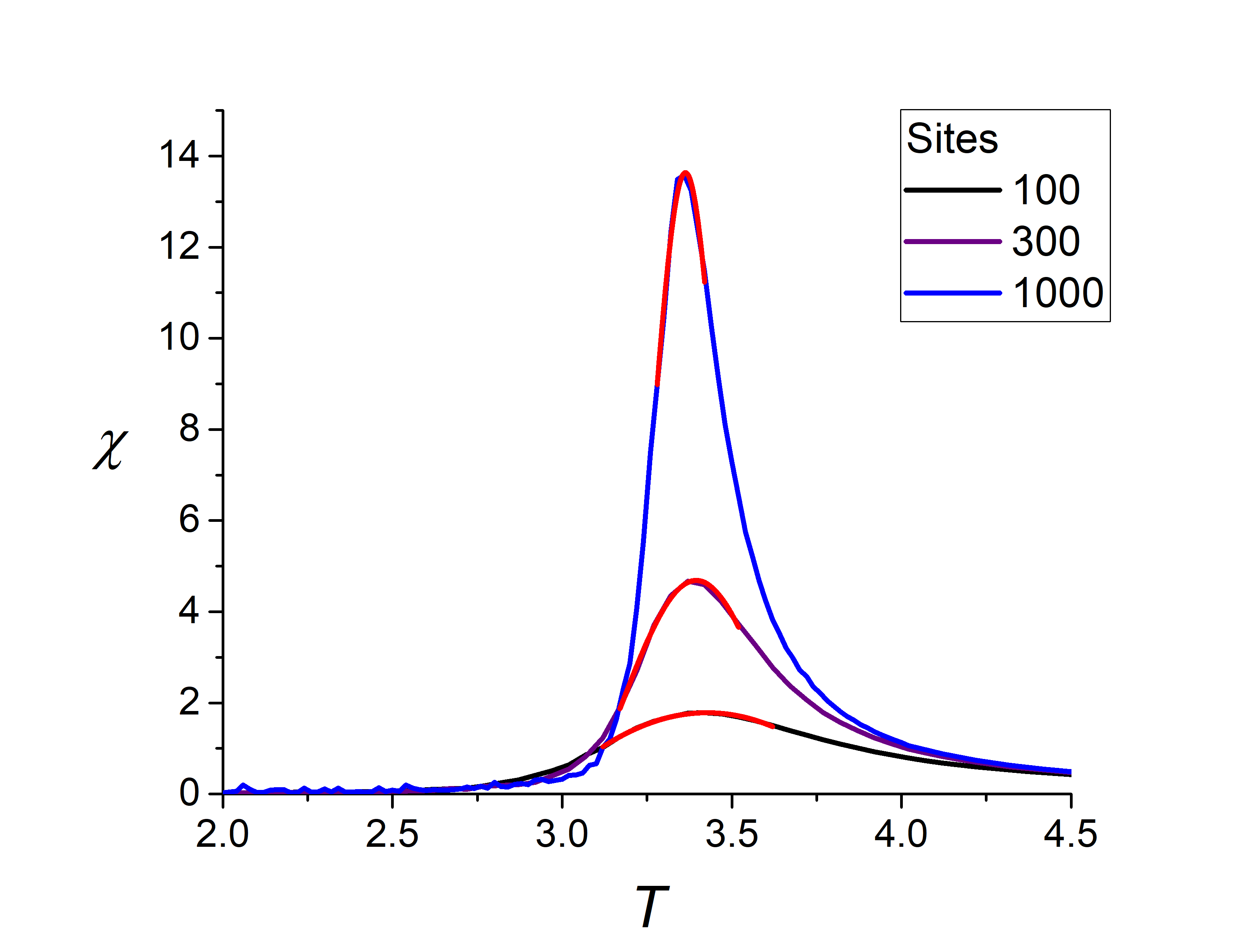}
\caption{(top) Heat capacity and (bottom) magnetic susceptibility for a Fibonacci sphere with $N = 100, 300, 1000$ sites. The red curves show the best fits by a Gaussian function and the region of the fits.}
\label{fig_chi_C}
\end{figure}

To illustrate the level of inconsistency further, we calculate the heat capacity 
\begin{equation}
C(T, N) = \frac{\langle E^2(T, N)\rangle - \langle E(T, N) \rangle^2}{T^2}\,.
\label{eq_C} 
\end{equation}
and the magnetic susceptibility
\begin{equation}
\chi(T, N) = \frac{\langle M^2(T, N)\rangle -\langle M(T, N) \rangle^2}{T}\,,
\label{eq_chi} 
\end{equation} 
in the pseudocritical region of couplings for a series of lattices with an increased total number of sites $N$. So far, these quantities have the standard behavior: their peaks with the height, which increases with rising $N$, can be fit by a Gaussian function. The examples of the data and the fits are shown in Fig.~\ref{fig_chi_C}.

\begin{figure}[!htb]
\includegraphics[width=0.45\textwidth]{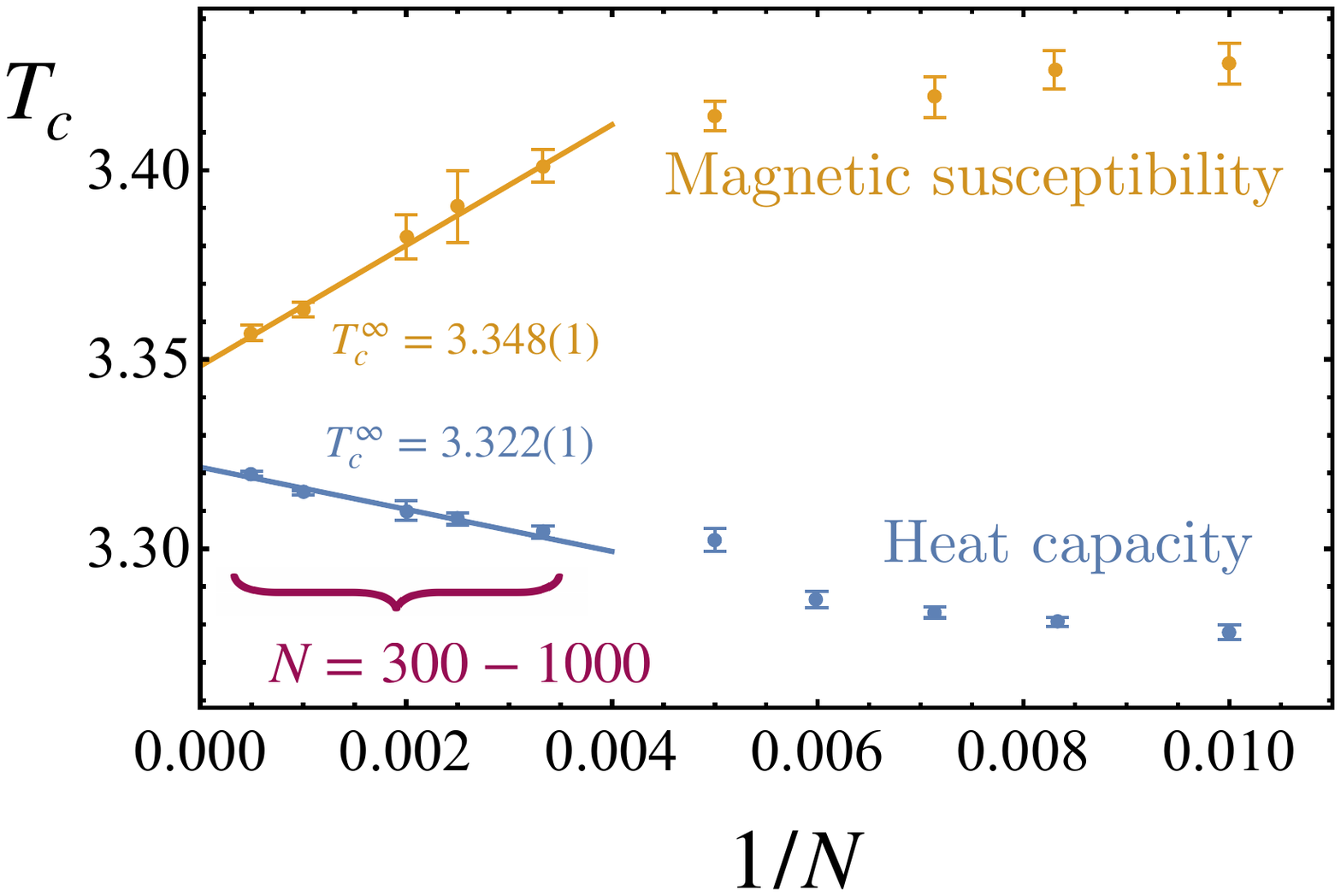}
\caption{The extrapolation of pseudocritical temperatures obtained from the heat capacity~\eq{eq_C}  and the magnetic susceptibility~\eq{eq_chi} towards the thermodynamic limit, $1/N \to 0$.}
\label{eq_extrapolation_Tc}
\end{figure}

The maxima of the fits point out to the pseudocritical temperatures, which are then plotted together as functions of an inverse number of sites of the Fibonacci sphere, $1/N$, in Fig.~\ref{eq_extrapolation_Tc}. The pseudocritical values, obtained from different quantities, do not coincide with each at finite $N$. According to our experience with planar lattices, this expected finite-volume effect should disappear in the thermodynamic limit. 

For the Fibonacci spheres, the pseudocritical temperatures can be extrapolated nicely towards the thermodynamic limit, but, surprisingly, the extrapolated values do not coincide with each other, Fig.~\ref{eq_extrapolation_Tc}. Although the difference between the extrapolated critical temperatures is about 1\%, the magnitude of this difference is substantial enough to be seen in our accurate calculations. This observation deepens the puzzle and gives us a quantitative counterpart of a similar effect that we observed on a qualitative level while studying the Binder cumulants.

\subsection{Thermodynamic limit}

\begin{figure}[!htb]
\includegraphics[width=0.45\textwidth]{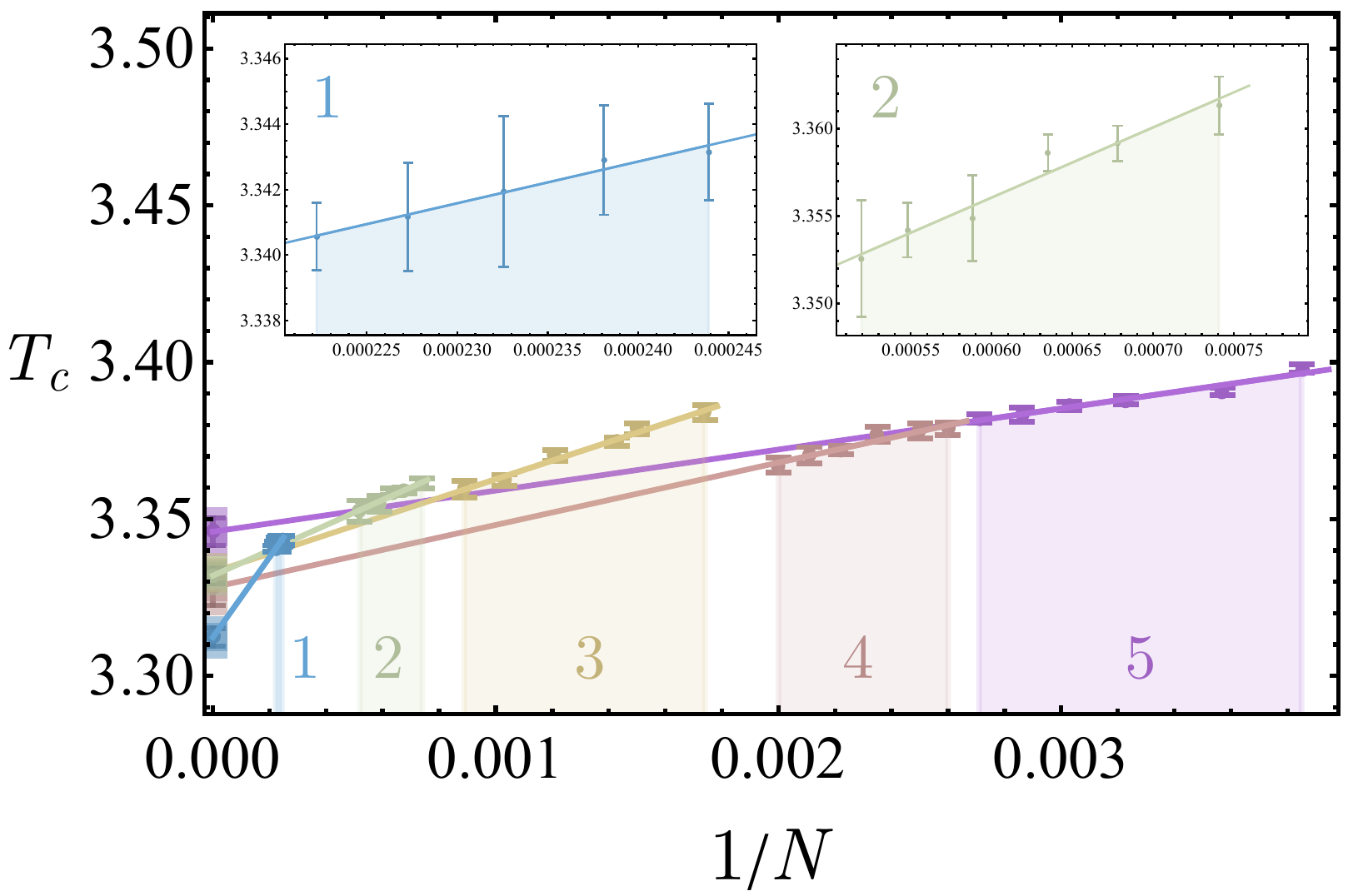}
\caption{The scaling of the pseudocritical temperature obtained from the magnetic susceptibility of the Ising model at the Fibonacci sphere as the function of an inverse number of sites, $1/N$. Five series of data, labeled $1,\dots,5$, are fitted separately by a linear dependence. The shaded regions at the horizontal axis mark the data ranges and the fits. The shaded regions at the vertical axis give the extrapolated values of the critical temperature and the corresponding extrapolation errors. The insets show zoom in on the regions 1 and 2.}
\label{fig_extrapolation}
\end{figure}

Does our result mean that the Fibonacci sphere, with the subsequent Delaunay triangulation, has no well-defined thermodynamic limit? That would be a surprising conclusion given that the Fibonacci construction provides us with the most uniform coverage of the sphere, which has no singular points and possesses other essential advantages with respect to the other triangulation schemes~\cite{gonzalez2010measurement}. To clarify this intriguing property, we plot in Fig.~\ref{fig_extrapolation} the extrapolation of the pseudocritical temperatures -- obtained from the magnetic susceptibility peaks -- towards the continuum limit using a more extensive set of Fibonacci spheres. 

Surprisingly, the data of Fig.~\ref{fig_extrapolation} shows that pseudo-critical temperatures can be separated into groups of different lattice sizes characterized by an almost perfect linear behavior. The slopes of the linear curves, however, differ from each other. Consequently, the extrapolated critical temperatures obtained from different size groups of the Fibonacci spheres give us slightly different critical temperatures in the continuum limit. 

This unusual phenomenon, the segmentation of the extrapolation region into separate intervals, can be related to sudden changes in the connectivity properties of the Fibonacci spheres. As we discuss in detail in Appendix~\ref{app_statistics}, the Fibonacci lattice consists of the sites connected to 5, 6, and 7 neighbors, with the 6-fold connectivity being dominant at large lattice sizes, $N \sim 10^3$. As we increase the lattice size, the proportion of the 6-connected sites increases up to a specific critical lattice size, $N = N^{(i)}_{\mathrm{cr}}$, when it abruptly diminishes singularly, and the rise of the fraction of the 6-connected sites starts over again. 

The critical values of the lattice size, $N = N^{(i)}_{\mathrm{cr}}$, correspond to the sudden restructuring of the connectivity properties of the Fibonacci sphere. In between the critical points, $(N^{(i)}_{\mathrm{cr}}, N^{(i+1)}_{\mathrm{cr}})$, the connectivity changes smoothly as the number of sites $N$ increases. The linear patches in Fig.~\ref{fig_extrapolation} correspond precisely to these smooth regions, while the change of the linear slopes occurs close to the critical value of $N$. 

The extrapolated critical values are shown in Fig.~\ref{fig_extrapolation}. The change in the connectivity properties of the Fibonacci lattice can be attributed to the systematic effect of the triangulation, which decreases as the lattice size increases. The combined region for extrapolated critical temperature thus gives us a consolidated error which includes statistical, fitting, and systematic effects. Our final result for the critical transition temperature of the Ising model on the Fibonacci sphere extrapolated to the thermodynamic limit is as follows:
\beqn
 \frac{T^{\large\ostar}_c}{J} \simeq 3.33(3) \simeq \frac{10}{3}\,,
\label{eq_T_c_F}
\eeqn
where ${\large\ostar}$ denotes the Delaunay-triangulated Fibonacci sphere.

Interestingly, the extrapolated critical temperature of the Ising model at the Fibonacci sphere lies in between the critical temperatures for the square lattice $T^\Box_c$ and that of the equilateral triangular lattice $T^\triangle_c$, which are given in Eqs.~\eq{eq_T_square} and \eq{eq_T_triangle}, respectively:
\beqn
T^\Box_c < T^{\large\ostar}_c \lesssim T^\triangle_c \,.
\eeqn
Naively, one could expect that $T^{\large\ostar}_c$ should coincide with $T^\triangle_c$ because both models are defined on the triangular lattices while the triangulated sphere in the thermodynamic limit $N \to \infty$ becomes a planar lattice. However, we cannot confirm this expectation with definiteness since in the large-$N$ limit, the Fibonacci lattice contains a significant (at a level of 1\%) portion of 5- and 7-connected sites as one can see in Fig.~\ref{fig_connectivity_lattice}. The critical temperature of the Ising model on the Fibonacci lattice is thus much closer to its triangular (6-connected) planar counterpart $T^\triangle_c$ than the critical temperature $T^\Box_c$ at the (4-connected) square lattice. The closeness of the two critical temperatures, $T^{\large\ostar}_c$ and $T^\triangle_c$, is well explained by the similarity of the connectivity properties of the Fibonacci triangulation in the large-$N$ limit and the planar equilateral triangle lattices. 

\section{Conclusions}
\label{sec:conclusions}

We studied the Ising model on a two-dimensional sphere triangulated using the Delaunay method applied to a uniform Fibonacci covering. This construction provides us with a uniform isotropic covering of the sphere with approximately equal-area triangles, thus potentially supporting a smooth thermodynamic limit. 

We found, however, that as the number of sites $N$ of the triangulated Fibonacci sphere increases, the connectivity properties of the triangular lattice experience discontinuous transitions at specific critical lattice sizes $N = N_{\mathrm{cr}}^{(i)}$, $i=1,2,\dots$ which influence its statistical features. While in the standard triangular lattice, every site is connected to 6 neighboring sites, the Fibonacci lattice contains a substantial density of the 5- and 7-vertices. The relative number of sites with different connectivity changes at critical sizes of the spheres, thus systematically affecting the extrapolation of the finite-volume results toward the thermodynamic limit. 

We uncovered that the Ising model on Fibonacci-triangulated spheres up to large sizes (with $2\times 10^5$ sites) possesses a single (pseudo)critical temperature associated with a spontaneous magnetization of the system. In the thermodynamic limit, the extrapolated critical temperature is given in Eq.~\eq{eq_T_c_F}. Its value slightly differs from the critical temperature of the Ising model on an equilateral triangular planar lattice~\eq{eq_T_triangle}. We interpret this result as a signature of a memory effect that can survive in the thermodynamic limit: the planar lattice corresponding to the Fibonacci sphere of an infinite radius remembers its curved origin imprinted in the irregular (but thermodynamically important) connectivity of the lattice sites.

\acknowledgments
The numerical calculations were performed on computing cluster Vostok-1 of Far Eastern Federal University. The work of AVM and ASP was supported by Grant No. 0657-2020-0015 of the Ministry of Science and Higher Education of Russia. 

\appendix
\section{Delaunay triangulation of a sphere}
\label{app:Delaunay}

A triangulation corresponds to a procedure that connects a set of sites (nodes) by non-intersecting links, thus covering the surface with a set of triangles. The Delaunay triangulation is a particular kind of procedure that is identified by the following criteria.

For a triangulation of a plane, the global criterion requires that a link belongs to the Delaunay triangulation if and only if one can represent this link as a chord of a circle that contains, in its interior, no other sites of the given set of sites. The local criterion demands that a link belongs to the Delaunay triangulation if and only if a circle that goes through all three vertices of any triangle does not contain, in its interior, the sites (nodes) of any adjacent triangle. 

For the Delaunay triangulation for a non-planar surface, for example, for a sphere, one can formulate the following condition that incorporates both the global and local criteria mentioned above. For every triangle of the set, we put in correspondence an auxiliary sphere such that this sphere incorporates all three nodes of the triangle while the sphere's center belongs to the plane of the triangle. For a correct Delaunay triangulation of the sphere, the interior of each auxiliary sphere does not contain any vertices of any other triangles. 

We implement the Delaunay triangulation using the following numerical algorithm:
\begin{enumerate}
    \item select a site $A$ from the set;

    \item take randomly two sites $B$ and $C$ from the same set as candidates for a possible triangle $ABC$ of the valid triangulation;
	
    \item verify whether any other site of the complete set belongs to the interior of the auxiliary sphere spanned on the triangle $ABC$ with the center located in the plane parallel to the plane of the $ABC$ triangle.
	
    \item If the response is negative, then the site $ABC$ belongs to the Delaunay triangulation. Otherwise, for site $A$, we choose other sites $B$ and $C$ from the set and repeat the procedure. 

\end{enumerate}

Step 3 is performed using the following construction. First, we need to find the position ${\bs x}_0 = (x_0, y_0, z_0)$  of the center of a circle described around a triangle and belonging to the triangle plane. 

The equation of a plane has the following form: 
\beqn
a x + b y + c z + d = 0,
\label{eq_plane_1}
\eeqn
where $a$, $b$, $c$, and $d$ are certain coefficients. Let the vertices of the triangle be given by the coordinates ${\bs x}_i =(x_i, y_i, z_i)$, where $i = 1,2$, and $3$ stand for the points $A$, $B$, and $C$, respectively. Then, the equation of the plane which passes through these three sites is defined by the following determinant:
\begin{equation}
\begin{vmatrix}
  x & y & z & 1\\
  x_1& y_1& z_1& 1\\
  x_2& y_2& z_2& 1\\
  x_3& y_3& z_3& 1\\
\end{vmatrix} = 0\,,
\label{eq_plane_2}
\end{equation}
which is equivalent to Eq.~\eq{eq_plane_1}. A comparison of Eqs.~\eq{eq_plane_1} and \eq{eq_plane_2} gives us the coefficients that determine the plane:
\begin{align}
a & = y_1 (z_2 - z_3) +y_2 (z_3 - z_1) + y_3 (z_1 - z_2),\\
b & = z_1 (x_2 - x_3) +z_2 (x_3 - x_1) + z_3 (x_1 - x_2),\\
b & = x_1 (y_2 - y_3) +x_2 (y_3 - y_1) + x_3 (y_1 - y_2),\\
d & = - \epsilon^{ijk} x_i y_j z_k\,,
\label{eq_coeff1}
\end{align}
where $\epsilon^{ijk}$ is Levi-Civita symbol.

The auxiliary sphere intersects the plane at a circle. The sphere and the circle share the same center ${\bs x}_0$, which is equally elongated from positions of the vertices ${\bs x}_i$, $i=1,2,3$ of the triangle. This requirement gives us the following three equations:
\begin{align}
    ({\bs x}_0 - {\bs x}_i)^2 = R^2\,, \qquad i = 1,2,3\,,
\label{eq_equally_spaced}
\end{align}
where 
\begin{equation}
R = \frac{AB\cdot BC\cdot CA}{4 S}\,,
\label{eq_R}
\end{equation}
is the radius of the circle that belongs to the $ABC$ plane and goes through the points $A$, $B$, and $C$. Here $AB$ is the length of the link $AB$ (and similarly for $BC$ and $CA$),
\begin{equation}\label{b} 
S \equiv S_{ABC} {=} \sqrt{p\cdot(p {-} AB)\cdot(p {-} BC)\cdot(p {-} CA)}\,.
 \end{equation}
is the area of the triangle and 
\begin{align}
    p \equiv p_{ABC} = \frac{AB + BC + CA}{2},
\end{align}
is its half-perimeter.

Setting ${\bs x} = {\bs x}_0$ in Eq.~\eq{eq_plane_1} and taking a difference between Eqs.~\eq{eq_equally_spaced} with $i=1$ and $i=2$, as well as $i=1$ and $i = 3$, we get three independent linear equations that can be combined into the single matrix equation of the following form:
\beqn
{\hat M} {\bs x}_0 + {\bs d} = 0\,,
\label{eq_linear}
\eeqn
with
\begin{align}
    {\hat M} = 
\begin{pmatrix}
  a & b & c\\
  a_1& b_1 & c_1\\
  a_2& b_2 & c_2\\
\end{pmatrix}\,, 
\qquad
d = 
\begin{pmatrix}
  d\\
  d_1\\
  d_2\\
\end{pmatrix}
\label{eq_M_d}
\end{align}
where
\begin{align}
a_1& = 2(x_2-x_1), \quad & a_2  = 2(x_3-x_1), \\
b_1 & =2(y_2-y_1), \quad & b_2  = 2(y_3-y_1), \\
b_1 & =2(z_2-z_1), \quad & c_2  = 2 (z_3-z_1), \\
d_1 & = ({\bs x_1})^2 - ({\bs x_2})^2, \quad   & \quad d_2 = ({\bs x_1})^2 - ({\bs x_3})^2.
\label{eq_coeff2}
\end{align}

The center of the circle is given by the solution of the linear equation~\eq{eq_linear}, ${\bs x}_0 = - {\hat M} {\bs d}$, for which all coefficients are known and given in Eqs.~\eq{eq_coeff1}, \eq{eq_M_d} and \eq{eq_coeff2}. The radius of the circle~\eq{eq_R} allows us to check the validity of the Delaunay triangulation, which implies that the inequality 
\beqn
({\bs x}_i - {\bs x}_0)^2 > R^2\,,
\eeqn
is satisfied for all points from the set other than the original $i=1,2,3$. Repeating this procedure for all points allows us to cover the sphere completely with triangular packages without empty areas and overlaps. 

Finally, to confirm that our triangulation does not produce artifacts such as intersections of links or incomplete coverage of the sphere with plaquettes, it is necessary to prove that the Fibonacci lattice topologically corresponds to the sphere. For this purpose, we calculate the Euler characteristic number $\chi = V - R + G$ of the discretized surface (here $V$, $R$, and $G$ are, respectively, the numbers of its vertices, edges, and faces of a polyhedron corresponding to the triangulation). Our calculation shows that the Euler characteristic for each triangulated Fibonacci lattice is $\chi = 2$, implying its topological equivalence to the sphere.

\begin{figure}[!htb]
\includegraphics[width=0.475\textwidth]{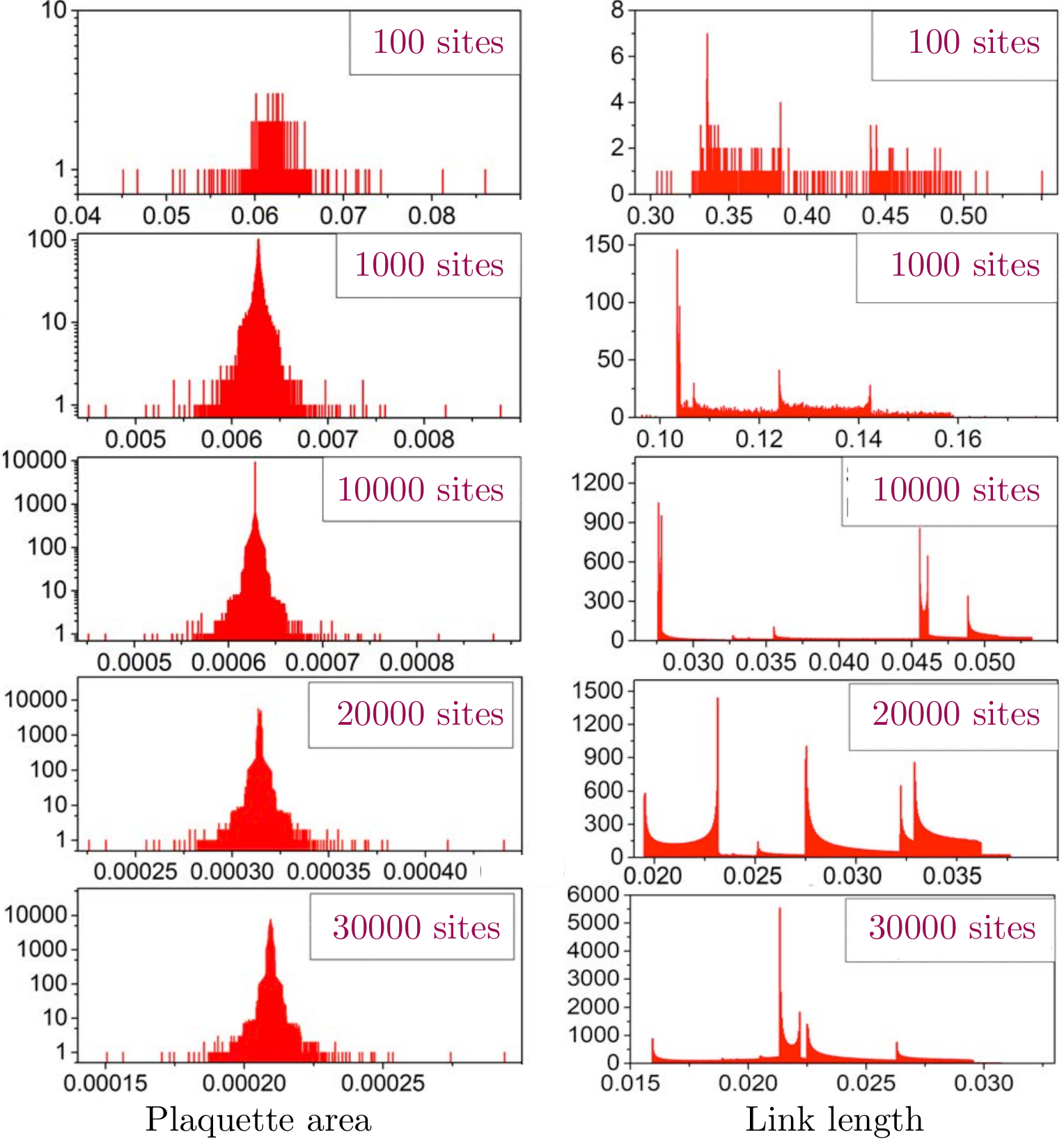}
\caption{Statistical distributions of the plaquette areas (left) and the link lengths (right) for the the Delaunay triangulation of the Fibonacci spheres of the unit radii for different total numbers of sites $N$. Notice the logarithmic scale of the area plots on the left.}
\label{fig_area_length_distributions}
\end{figure}

\begin{figure}[!htb]
\includegraphics[width=0.425\textwidth]{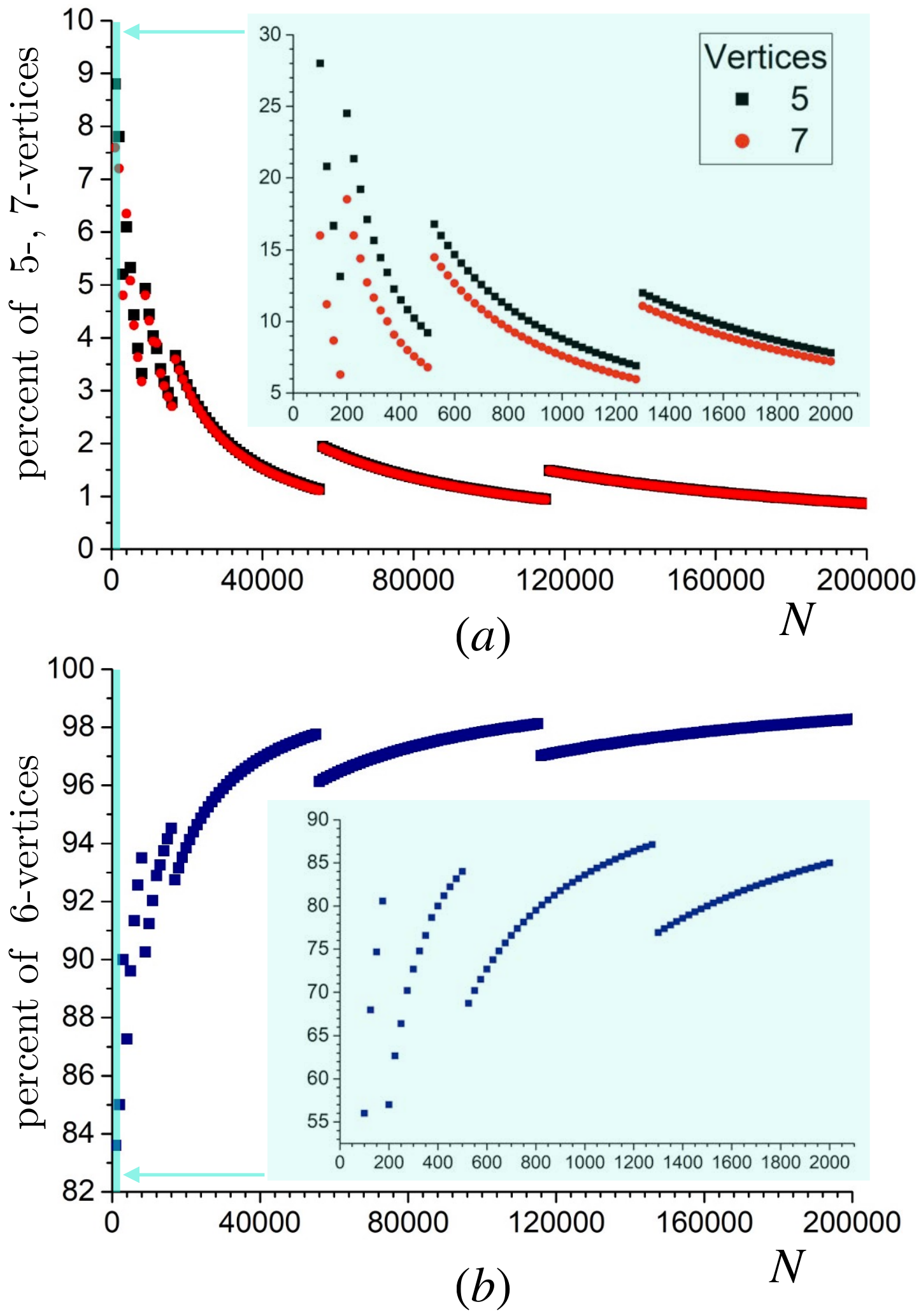}
\caption{The statistical connectivity properties of the Fibonacci spheres with the Delaunay triangulation: the percentage of (a) 5- and 7-vertex sites and (b) 6-vertex sites for lattices with a different number of sites $N$. The inset shows the region with a relatively small number of points (shown in the light blue color).}
\label{fig_connectivity_lattice}
\end{figure}

\section{Statistical properties of Delaunay triangulation of Fibonacci sphere}
\label{app_statistics}

One of the main advantages of the Fibonacci lattice is the partitioning of the surface into triangular plaquettes of approximately equal areas. To verify this property for the Delaunay triangulation of the sphere used in our work, we plot in Fig.~\ref{fig_area_length_distributions}(left) the distribution of plaquettes for the Delaunay triangulation for lattices of various sizes $N$. The plots show that our construction is characterized by a remarkably narrow distribution of the area of triangular plaquettes, which keeps its narrowness with increasing lattice size $N$.

However, the area distribution is not smooth: instead of the expected Gaussian profile, the area distribution of Fig.~\ref{fig_area_length_distributions}(left) shows an Empire State Building-like profile with several clearly visible steps. Such steps are probably related to inhomogeneous profiles of the statistical distributions of the length of the lattice links shown in Fig.~\ref{fig_area_length_distributions}(right). Thus, the approximate equality of the areas of the triangular plaquettes does not imply an equality of the lengths of the links connecting neighboring lattice nodes. In other words, the statistical distribution of link lengths is not characterized by a narrow distribution. However, this fact does not affect our construction because the Ising Hamiltonian~\eq{eq_H} does not consider the length of the links between adjacent sites. 

Another attractive characteristic of the Fibonacci sphere is the connectivity of the lattice, which does affect the Ising Hamiltonian~\eq{eq_H}. We already mentioned in the Introduction that the physical properties of the Ising model defined on planar lattices with different connectivities (for example, square and triangular lattices) differ. Each site of the square lattice possesses (is connected with) four nearest neighbors, while the number of neighbors of the equilateral triangular lattice is six. The number of nearest neighbors of the Fibonacci lattice sites varies from 5 to 7, as illustrated in Fig.~\ref{fig_Delaunay}(b). The most common number of neighbors (the ``connectivity'') is 6. 

The significant heterogeneity of neighbors affects the statistical properties of the Ising model at Fibonacci lattices. The statistical connectivity properties of the Fibonacci spheres with the Delaunay triangulation are shown in Fig.~\ref{fig_connectivity_lattice}. One can observe that the percentage of such 6-connected sites generally increases as the number of sites increases. However, the uniform monotonic increase gets broken in isolated critical points $N^{(i)}_{\mathrm{cr}}$, $i=1,2,\dots$ where the number of sites with the 6-point connectivity experiences sudden jumps towards lower values. At the same time, the number of sites with 5- and 7-point connectivities shows discontinuities towards higher values at the same set of critical points. We observed that the jumps are related to the formation of new Fibonacci spirals, Fig.~\ref{fig_spirals}, in the covering of the sphere.

The discontinuities in connectivity are pronounced at small sizes of the Fibonacci sphere with $O(30\%)$ jumps at $N \sim 10^2$ and $O(10\%)$ discontinuities at $N \sim 10^3$, while at larger lattices, $N \sim 10^5$, the jumps become much smoother, reducing to the $O(1\%)$ order of magnitude.

\bibliographystyle{apsrev4-1}

\begin{thebibliography}{22}%
\makeatletter
\providecommand \@ifxundefined [1]{%
 \@ifx{#1\undefined}
}%
\providecommand \@ifnum [1]{%
 \ifnum #1\expandafter \@firstoftwo
 \else \expandafter \@secondoftwo
 \fi
}%
\providecommand \@ifx [1]{%
 \ifx #1\expandafter \@firstoftwo
 \else \expandafter \@secondoftwo
 \fi
}%
\providecommand \natexlab [1]{#1}%
\providecommand \enquote  [1]{``#1''}%
\providecommand \bibnamefont  [1]{#1}%
\providecommand \bibfnamefont [1]{#1}%
\providecommand \citenamefont [1]{#1}%
\providecommand \href@noop [0]{\@secondoftwo}%
\providecommand \href [0]{\begingroup \@sanitize@url \@href}%
\providecommand \@href[1]{\@@startlink{#1}\@@href}%
\providecommand \@@href[1]{\endgroup#1\@@endlink}%
\providecommand \@sanitize@url [0]{\catcode `\\12\catcode `\$12\catcode
  `\&12\catcode `\#12\catcode `\^12\catcode `\_12\catcode `\%12\relax}%
\providecommand \@@startlink[1]{}%
\providecommand \@@endlink[0]{}%
\providecommand \url  [0]{\begingroup\@sanitize@url \@url }%
\providecommand \@url [1]{\endgroup\@href {#1}{\urlprefix }}%
\providecommand \urlprefix  [0]{URL }%
\providecommand \Eprint [0]{\href }%
\providecommand \doibase [0]{http://dx.doi.org/}%
\providecommand \selectlanguage [0]{\@gobble}%
\providecommand \bibinfo  [0]{\@secondoftwo}%
\providecommand \bibfield  [0]{\@secondoftwo}%
\providecommand \translation [1]{[#1]}%
\providecommand \BibitemOpen [0]{}%
\providecommand \bibitemStop [0]{}%
\providecommand \bibitemNoStop [0]{.\EOS\space}%
\providecommand \EOS [0]{\spacefactor3000\relax}%
\providecommand \BibitemShut  [1]{\csname bibitem#1\endcsname}%
\let\auto@bib@innerbib\@empty
\bibitem [{\citenamefont {Landau}\ and\ \citenamefont {Lifshitz}(2013)}]{LL5}%
  \BibitemOpen
  \bibfield  {author} {\bibinfo {author} {\bibfnamefont {L.~D.}\ \bibnamefont
  {Landau}}\ and\ \bibinfo {author} {\bibfnamefont {E.~M.}\ \bibnamefont
  {Lifshitz}},\ }\href@noop {} {\emph {\bibinfo {title} {Statistical Physics:
  Volume 5}}},\ Vol.~\bibinfo {volume} {5}\ (\bibinfo  {publisher} {Elsevier},\
  \bibinfo {year} {2013})\BibitemShut {NoStop}%
\bibitem [{\citenamefont {Kramers}\ and\ \citenamefont
  {Wannier}(1941)}]{Kramers:1941kn}%
  \BibitemOpen
  \bibfield  {author} {\bibinfo {author} {\bibfnamefont {H.~A.}\ \bibnamefont
  {Kramers}}\ and\ \bibinfo {author} {\bibfnamefont {G.~H.}\ \bibnamefont
  {Wannier}},\ }\href {\doibase 10.1103/PhysRev.60.252} {\bibfield  {journal}
  {\bibinfo  {journal} {Phys. Rev.}\ }\textbf {\bibinfo {volume} {60}},\
  \bibinfo {pages} {252} (\bibinfo {year} {1941})}\BibitemShut {NoStop}%
\bibitem [{\citenamefont {Baxter}(2016)}]{baxter2016exactly}%
  \BibitemOpen
  \bibfield  {author} {\bibinfo {author} {\bibfnamefont {R.~J.}\ \bibnamefont
  {Baxter}},\ }\href@noop {} {\emph {\bibinfo {title} {Exactly solved models in
  statistical mechanics}}}\ (\bibinfo  {publisher} {Elsevier},\ \bibinfo {year}
  {2016})\BibitemShut {NoStop}%
\bibitem [{\citenamefont {Newell}(1950)}]{Newell_1950}%
  \BibitemOpen
  \bibfield  {author} {\bibinfo {author} {\bibfnamefont {G.~F.}\ \bibnamefont
  {Newell}},\ }\href {\doibase doi:10.1103/physrev.79.876 } {\bibfield
  {journal} {\bibinfo  {journal} {Physical Review}\ }\textbf {\bibinfo {volume}
  {79}},\ \bibinfo {pages} {876} (\bibinfo {year} {1950})}\BibitemShut
  {NoStop}%
\bibitem [{\citenamefont {Wannier}(1950)}]{Wannier_1950}%
  \BibitemOpen
  \bibfield  {author} {\bibinfo {author} {\bibfnamefont {G.~H.}\ \bibnamefont
  {Wannier}},\ }\href {https://arxiv.org/abs/2109.00254} {\bibfield  {journal}
  {\bibinfo  {journal} {Physical Review}\ }\textbf {\bibinfo {volume} {79}},\
  \bibinfo {pages} {357} (\bibinfo {year} {1950})}\BibitemShut {NoStop}%
\bibitem [{\citenamefont {Zhi-Huan}\ \emph {et~al.}(2009)\citenamefont
  {Zhi-Huan}, \citenamefont {Mushtaq}, \citenamefont {Yan},\ and\ \citenamefont
  {Jian-Rong}}]{Zhi_Huan_2009}%
  \BibitemOpen
  \bibfield  {author} {\bibinfo {author} {\bibfnamefont {L.}~\bibnamefont
  {Zhi-Huan}}, \bibinfo {author} {\bibfnamefont {L.}~\bibnamefont {Mushtaq}},
  \bibinfo {author} {\bibfnamefont {L.}~\bibnamefont {Yan}}, \ and\ \bibinfo
  {author} {\bibfnamefont {L.}~\bibnamefont {Jian-Rong}},\ }\href {\doibase
  doi:10.1088/1674-1056/18/7/012} {\bibfield  {journal} {\bibinfo  {journal}
  {Chinese Physics B}\ }\textbf {\bibinfo {volume} {18}},\ \bibinfo {pages}
  {2696} (\bibinfo {year} {2009})}\BibitemShut {NoStop}%
\bibitem [{\citenamefont {Collins}\ and\ \citenamefont
  {Petrenko}(1997)}]{collins1997review}%
  \BibitemOpen
  \bibfield  {author} {\bibinfo {author} {\bibfnamefont {M.}~\bibnamefont
  {Collins}}\ and\ \bibinfo {author} {\bibfnamefont {O.}~\bibnamefont
  {Petrenko}},\ }\href@noop {} {\bibfield  {journal} {\bibinfo  {journal}
  {Canadian journal of physics}\ }\textbf {\bibinfo {volume} {75}},\ \bibinfo
  {pages} {605} (\bibinfo {year} {1997})}\BibitemShut {NoStop}%
\bibitem [{\citenamefont {Gonz{\'a}lez}(2010)}]{gonzalez2010measurement}%
  \BibitemOpen
  \bibfield  {author} {\bibinfo {author} {\bibfnamefont {{\'A}.}~\bibnamefont
  {Gonz{\'a}lez}},\ }\href@noop {} {\bibfield  {journal} {\bibinfo  {journal}
  {Mathematical Geosciences}\ }\textbf {\bibinfo {volume} {42}},\ \bibinfo
  {pages} {49} (\bibinfo {year} {2010})}\BibitemShut {NoStop}%
\bibitem [{\citenamefont {Saliba}\ and\ \citenamefont
  {Barnes}(2019)}]{Saliba_2019}%
  \BibitemOpen
  \bibfield  {author} {\bibinfo {author} {\bibfnamefont {E.~P.}\ \bibnamefont
  {Saliba}}\ and\ \bibinfo {author} {\bibfnamefont {A.~B.}\ \bibnamefont
  {Barnes}},\ }\href {\doibase https://doi.org/10.1063/1.5113598} {\bibfield
  {journal} {\bibinfo  {journal} {The Journal of Chemical Physics}\ }\textbf
  {\bibinfo {volume} {151}},\ \bibinfo {pages} {114107} (\bibinfo {year}
  {2019})}\BibitemShut {NoStop}%
\bibitem [{\citenamefont {Marques}\ \emph {et~al.}(2013)\citenamefont
  {Marques}, \citenamefont {Bouville}, \citenamefont {Ribardi{\`{e}}re},
  \citenamefont {Santos},\ and\ \citenamefont {Bouatouch}}]{Marques_2013}%
  \BibitemOpen
  \bibfield  {author} {\bibinfo {author} {\bibfnamefont {R.}~\bibnamefont
  {Marques}}, \bibinfo {author} {\bibfnamefont {C.}~\bibnamefont {Bouville}},
  \bibinfo {author} {\bibfnamefont {M.}~\bibnamefont {Ribardi{\`{e}}re}},
  \bibinfo {author} {\bibfnamefont {L.~P.}\ \bibnamefont {Santos}}, \ and\
  \bibinfo {author} {\bibfnamefont {K.}~\bibnamefont {Bouatouch}},\ }\href
  {\doibase 10.1111/cgf.12190} {\bibfield  {journal} {\bibinfo  {journal}
  {Computer Graphics Forum}\ }\textbf {\bibinfo {volume} {32}},\ \bibinfo
  {pages} {134} (\bibinfo {year} {2013})}\BibitemShut {NoStop}%
\bibitem [{\citenamefont {Swinbank}\ and\ \citenamefont
  {Purser}(2006)}]{Swinbank_2006}%
  \BibitemOpen
  \bibfield  {author} {\bibinfo {author} {\bibfnamefont {R.}~\bibnamefont
  {Swinbank}}\ and\ \bibinfo {author} {\bibfnamefont {R.~J.}\ \bibnamefont
  {Purser}},\ }\href {\doibase https://doi.org/10.1256/qj.05.227} {\bibfield
  {journal} {\bibinfo  {journal} {Quarterly Journal of the Royal Meteorological
  Society}\ }\textbf {\bibinfo {volume} {132}},\ \bibinfo {pages} {1769}
  (\bibinfo {year} {2006})}\BibitemShut {NoStop}%
\bibitem [{\citenamefont {Song}\ \emph {et~al.}(2022)\citenamefont {Song},
  \citenamefont {Gao}, \citenamefont {Hou}, \citenamefont {Wang}, \citenamefont
  {Zhou}, \citenamefont {He}, \citenamefont {Guo},\ and\ \citenamefont
  {Chien}}]{Song2021}%
  \BibitemOpen
  \bibfield  {author} {\bibinfo {author} {\bibfnamefont {C.-H.}\ \bibnamefont
  {Song}}, \bibinfo {author} {\bibfnamefont {Q.-C.}\ \bibnamefont {Gao}},
  \bibinfo {author} {\bibfnamefont {X.-Y.}\ \bibnamefont {Hou}}, \bibinfo
  {author} {\bibfnamefont {X.}~\bibnamefont {Wang}}, \bibinfo {author}
  {\bibfnamefont {Z.}~\bibnamefont {Zhou}}, \bibinfo {author} {\bibfnamefont
  {Y.}~\bibnamefont {He}}, \bibinfo {author} {\bibfnamefont {H.}~\bibnamefont
  {Guo}}, \ and\ \bibinfo {author} {\bibfnamefont {C.-C.}\ \bibnamefont
  {Chien}},\ }\href@noop {} {\bibfield  {journal} {\bibinfo  {journal}
  {Physical Review Research}\ }\textbf {\bibinfo {volume} {4}},\ \bibinfo
  {pages} {023005} (\bibinfo {year} {2022})}\BibitemShut {NoStop}%
\bibitem [{\citenamefont {Saff}\ and\ \citenamefont
  {Kuijlaars}(1997)}]{Saff_1997}%
  \BibitemOpen
  \bibfield  {author} {\bibinfo {author} {\bibfnamefont {E.~B.}\ \bibnamefont
  {Saff}}\ and\ \bibinfo {author} {\bibfnamefont {A.~B.~J.}\ \bibnamefont
  {Kuijlaars}},\ }\href {\doibase https://doi.org/10.1007/BF03024331}
  {\bibfield  {journal} {\bibinfo  {journal} {The Mathematical Intelligencer}\
  }\textbf {\bibinfo {volume} {19}},\ \bibinfo {pages} {5} (\bibinfo {year}
  {1997})}\BibitemShut {NoStop}%
\bibitem [{\citenamefont {Niederreiter}\ and\ \citenamefont
  {H.}(1994)}]{Niederreiter_1994}%
  \BibitemOpen
  \bibfield  {author} {\bibinfo {author} {\bibfnamefont {H.}~\bibnamefont
  {Niederreiter}}\ and\ \bibinfo {author} {\bibfnamefont {S.~I.}\ \bibnamefont
  {H.}},\ }\href {\doibase doi:10.1016/0377-0427(92)00004-s} {\bibfield
  {journal} {\bibinfo  {journal} {Journal of Computational and Applied
  Mathematics}\ }\textbf {\bibinfo {volume} {51}},\ \bibinfo {pages} {57}
  (\bibinfo {year} {1994})}\BibitemShut {NoStop}%
\bibitem [{\citenamefont {Vogel}(1979)}]{Vogel_1979}%
  \BibitemOpen
  \bibfield  {author} {\bibinfo {author} {\bibfnamefont {H.}~\bibnamefont
  {Vogel}},\ }\href {\doibase doi:10.1016/0025-5564(79)90080-4} {\bibfield
  {journal} {\bibinfo  {journal} {Mathematical Biosciences}\ }\textbf {\bibinfo
  {volume} {44}},\ \bibinfo {pages} {179} (\bibinfo {year} {1979})}\BibitemShut
  {NoStop}%
\bibitem [{\citenamefont {Bravais}\ and\ \citenamefont
  {Bravais}(1837)}]{Bravais1837}%
  \BibitemOpen
  \bibfield  {author} {\bibinfo {author} {\bibfnamefont {L.}~\bibnamefont
  {Bravais}}\ and\ \bibinfo {author} {\bibfnamefont {A.}~\bibnamefont
  {Bravais}},\ }\href@noop {} {\bibfield  {journal} {\bibinfo  {journal} {Ann.
  Sci. Nat.Bot. Biol. Vég.}\ }\textbf {\bibinfo {volume} {7}},\ \bibinfo
  {pages} {11–42, 42–110, 193–221, 291–348.} (\bibinfo {year}
  {1837})}\BibitemShut {NoStop}%
\bibitem [{\citenamefont {Weisstein}(2002)}]{weisstein2002crc}%
  \BibitemOpen
  \bibfield  {author} {\bibinfo {author} {\bibfnamefont {E.~W.}\ \bibnamefont
  {Weisstein}},\ }\href@noop {} {\emph {\bibinfo {title} {CRC concise
  encyclopedia of mathematics}}}\ (\bibinfo  {publisher} {Chapman and
  Hall/CRC},\ \bibinfo {year} {2002})\BibitemShut {NoStop}%
\bibitem [{\citenamefont {Gonz{\'{a}}lez}(2009)}]{Gonz_lez_2009}%
  \BibitemOpen
  \bibfield  {author} {\bibinfo {author} {\bibfnamefont {{\'{A}}.}~\bibnamefont
  {Gonz{\'{a}}lez}},\ }\href {\doibase 10.1007/s11004-009-9257-x} {\bibfield
  {journal} {\bibinfo  {journal} {Mathematical Geosciences}\ }\textbf {\bibinfo
  {volume} {42}},\ \bibinfo {pages} {49} (\bibinfo {year} {2009})}\BibitemShut
  {NoStop}%
\bibitem [{\citenamefont {Hannay}\ and\ \citenamefont
  {Nye}(2004)}]{Hannay_2004}%
  \BibitemOpen
  \bibfield  {author} {\bibinfo {author} {\bibfnamefont {J.~H.}\ \bibnamefont
  {Hannay}}\ and\ \bibinfo {author} {\bibfnamefont {J.~F.}\ \bibnamefont
  {Nye}},\ }\href {\doibase doi:10.1088/0305-4470/37/48/005} {\bibfield
  {journal} {\bibinfo  {journal} {Journal of Physics A: Mathematical and
  General}\ }\textbf {\bibinfo {volume} {37}},\ \bibinfo {pages} {11591}
  (\bibinfo {year} {2004})}\BibitemShut {NoStop}%
\bibitem [{\citenamefont {Keinert}\ \emph {et~al.}(2015)\citenamefont
  {Keinert}, \citenamefont {Innmann}, \citenamefont {Sänger},\ and\
  \citenamefont {Stamminger}}]{Keinert_2015}%
  \BibitemOpen
  \bibfield  {author} {\bibinfo {author} {\bibfnamefont {B.}~\bibnamefont
  {Keinert}}, \bibinfo {author} {\bibfnamefont {M.}~\bibnamefont {Innmann}},
  \bibinfo {author} {\bibfnamefont {M.}~\bibnamefont {Sänger}}, \ and\
  \bibinfo {author} {\bibfnamefont {M.}~\bibnamefont {Stamminger}},\ }\href
  {\doibase 10.1145/2816795.2818131} {\bibfield  {journal} {\bibinfo  {journal}
  {{ACM} Transactions on Graphics}\ }\textbf {\bibinfo {volume} {34}},\
  \bibinfo {pages} {1} (\bibinfo {year} {2015})}\BibitemShut {NoStop}%
\bibitem [{\citenamefont {Delaunay}(1934)}]{Delaunay1934}%
  \BibitemOpen
  \bibfield  {author} {\bibinfo {author} {\bibfnamefont {B.}~\bibnamefont
  {Delaunay}},\ }\href@noop {} {\bibfield  {journal} {\bibinfo  {journal}
  {Bulletin de l’Académie des Sciences de l’URSS. Classe des sciences
  mathématiques et naturelles}\ }\textbf {\bibinfo {volume} {6}},\ \bibinfo
  {pages} {793–800} (\bibinfo {year} {1934})}\BibitemShut {NoStop}%
\bibitem [{\citenamefont {Skvortsov}(2002)}]{Skvortsov2002}%
  \BibitemOpen
  \bibfield  {author} {\bibinfo {author} {\bibfnamefont {A.}~\bibnamefont
  {Skvortsov}},\ }\href@noop {} {\bibfield  {journal} {\bibinfo  {journal}
  {Publishing House of Tomsk State University}\ ,\ \bibinfo {pages} {128}}
  (\bibinfo {year} {2002})}\BibitemShut {NoStop}%
\end{thebibliography}

%

\end{document}